\documentclass[twocolumn,showpacs,preprintnumbers,amsmath,amssymb]{revtex4}
\usepackage{dcolumn}
\usepackage{bm}
\usepackage{epsfig}

\usepackage{dcolumn}
\usepackage{bm}
\usepackage{epsfig}

\newcommand{\pa}{\partial}

\renewcommand{\bar}[1]{\overline{#1}}

\usepackage{amssymb}

\renewcommand{\bar}[1]{\overline{#1}}

\usepackage{indentfirst}
\usepackage{psfig,color}
\usepackage{epsfig}
\usepackage{graphicx}

\providecommand{\Journal}[4] {#1 {\bf #2} (#4) #3}
\providecommand{\NPA}{Nucl. Phys. A } %
\providecommand{\NPB}{Nucl. Phys. B } %
\providecommand{\PLB}{Phys. Lett. B } %
\providecommand{\PRL}{Phys. Rev. Lett. } %
\providecommand{\PRC}{Phys. Rev. C } %
\providecommand{\PRD}{Phys. Rev. D } %
\providecommand{\PRSA}{Proc. Roy. Soc. A } %
\providecommand{\RMP}{Rev. Mod. Phys. } %
\providecommand{\RMP}{Rev. Mod. Phys. } %
\providecommand{\ZPA}{Z. Phys. A } %

\begin{document}

\title{X(1835) as Proton-Antiproton Bound State in Skyrme Model\footnote{Talk presented in
{\it the 3rd meeting on the new BES physics}, Nanning, Nov., 2005.
}}

\author{Mu-Lin Yan\footnote{email:mlyan@ustc.edu.cn}}
\affiliation{Interdisciplinary Center for Theoretical Study,
Department of Modern Physics,\\
 University of Science and Technology
of China, Hefei, Anhui 230026, China}


\begin{abstract}
We present a review to the recent works related to interpreting the
exotic particle $X(1835)$ reported by BES as a  $N\bar{N}$-baryonium
in the Skyrme model.
  There are two  evidences that support this interpretation: 1) There exist a classical
$N\bar{N}$-Skyrmion solution with about $\sim 10$MeV binding
energies in the model; 2) The decay of this Skymion-baryonium is
caused by annihilation of $p-\bar{p}$ inside $X(1835)$ through the
quantum tunneling effect, and hence the most favorable decay
channels are $X\rightarrow \eta 4\pi$ or $X\rightarrow \eta' 2\pi$.
These lead to  reasonable interpreting  the data of BES, and
especially to useful prediction on the decay mode of $X(1835)$ for
the experiment. \vskip0.2in

\end{abstract}


\maketitle

\section{Introduction}
\noindent By using a sample of $5.8\times 10^7 J/\psi$ events
collected with the BES II detector,   BES collaboration studied
$J/\psi\rightarrow \gamma p\bar{p}$ at 2003, and found out an
anomalous enhancement near the mass threshold in the $p\bar{p}$
invariant-mass spectrum from this  decay process\cite{BES}. This
enhancement was fitted with a subthreshold $S$-wave Breit-Wigner
(BW) resonance function with a mass $M=1859^{+3+5}_{-10-25} {\rm
MeV}/c^2$, a width $\Gamma <30 {\rm MeV}/c^2$ (at the $90\%$ C.L.),
and a product branching fraction (BF) $B(J/\psi \rightarrow \gamma
X)\cdot B(X \rightarrow p\bar{p})=[7.0\pm
0.4(stat)^{+1.9}_{-0.8}(syst)]\times 10^{-5}$. These observations
could be naively interpreted as signals for baryonium $p\bar{p}$
bound state\cite{FY,Datta,Yan1,Yan2,Yan3,Yan4, GZh}, and will be
denoted as $X(1835)$ hereafter. The BES-datum fit in Ref.\cite{BES}
represents the simplest interpretation of the experimental results
as indication of a baryonium resonance. As an unstable particle, the
decays of $X(1835)$ must be caused by the proton-antiproton
annihilation inside the bound state $X(1835)$\cite{Yan1,Yan2}. This
should be regarded as a significant feature for distinguishing
$X(1835)$'s baryonium interpretation from other ones, say glueball,
hybrid, or $\eta'$'s excitation etc. In Ref.\cite{Yan1}, the $p
\bar{p}$-system has been studied by means of the Skyrme model. And
an attractive potential at middle distance scale range between $p$
and $\bar{p}$, and a repulsive force at short distance of
$p$-$\bar{p}$ has been revealed. This means $X(1835)$ can indeed be
understood as a baryonium in the Skyrmion framework. Thus, to
discuss the $p-\bar{p}$ annihilations insides $X(1835)$ in Skyrme
model is legitimate. In Ref.\cite{Yan2},  this issue  has been
investigated in detail by using coherent state method in the model
following the Amado-Cannata-Dedoder-Locher-Lu's studies to the
annihilations of $p-\bar{p}$ scattering\cite{batch1,Yang
lu3,batch2}. In this way, we found that $B(X\rightarrow \eta 4\pi)>>
B(X\rightarrow \eta 2\pi)$, and then we argue $B(X\rightarrow \eta'
2\pi)$ (with $\eta'\rightarrow \eta 2\pi$) must be much bigger than
$B(X\rightarrow \eta 2\pi)$. This unusual prediction provides a
criteria to identify whether $X(1835)$ is a baryonium or not.
Furthermore, to search the resonance of $\eta 4\pi$ or $\eta' 2\pi$
itself at the final state invariant mass $ 1800 \sim 1900$MeV may
reveal a full resonance peak for $X$-particle. Obviously, it is very
significant to show the resonance via $X$-particle's mesonic decays
both because the enhancement near the mass threshold in the
$p\bar{p}$ invariant-mass spectrum from $J/\psi\rightarrow \gamma
p\bar{p}$ decays has only shown a tail effect of the resonance and
because a narrow  resonance of $X\rightarrow \eta 2\pi$ has never
been seen. Actually, that there is no clear signal of narrow
resonance of $ \eta 2\pi$  at $ 1800 \sim 1900$MeV corresponding to
$X$-particle being seen in BES II was a serious obstacle to
understand the existence of $X$-particle at one time, because  the
quantum number assessment in Refs\cite{Datta,GZh} means
$X\rightarrow \eta\pi\pi$ should be the simplest decay mode for
$X$-particle with the largest phase space. In this case, then,
according to the analysis of Ref.\cite{Yan2}, to search the mesonic
resonance of $\eta 4\pi$ or $\eta' 2\pi$ with $ (\eta'\rightarrow
\eta 2\pi)$ was urged. Consequently, by searching these decay
processes, this obstacle was finely gotten over by a very beautiful
experiment of BES II\cite{BES2}. The full resonance peak of
$X(1835)$ has been revealed in the $J/\psi\rightarrow \gamma \eta'
\pi^+\pi^-$ channel with a statistical significance of $7.7\sigma
$\cite{BES2}. The $\eta'$ meson was detected in both $\eta \pi\pi$
and $\gamma \rho$ channels. There are roughly $264\pm 54$ events.
Its mass is $M_X=(1833.7\pm 6.2\pm 2.7)$MeV and its width is
$\Gamma(X(1835))=(67.7\pm20.3\pm7.7)$MeV\cite{BES2}. The mass and
width of the $X(1835)$ are not compatible with any known meson
resonance\cite{PDG}. In additional, the $S$-wave BW fit to the
$p\bar{p}$ invariant-mass spectrum of Ref.\cite{BES} was improved in
Ref.\cite{BES2} by further considering the final state interaction
effects in $X\rightarrow p\bar{p}$\cite{FSI,Sib}. The redoing
corrected the original results ($M \sim 1859{\rm MeV}/c^2$ with
$\Gamma <30 {\rm MeV}/c^2$) to be $M=1831\pm 7{\rm MeV}/c^2$ with
$\Gamma <153 {\rm MeV}/c^2$ (at the $90\%$ C.L.), which is
consistent with ones observed in Ref.\cite{BES2} from $X\rightarrow
\eta'\pi\pi$.  This is a strong evident to support that the
$p\bar{p}$-enhancement reported in Ref.\cite{BES} and the resonance
of $\eta'\pi\pi$ reported by Ref.\cite{BES2} are due to same
baryonium particle with $I^G(J^{PC})=0^+(0^{-+})$. Moreover, it has
been revealed also that the relative $p\bar{p}$ decay strength is
quite strong: $B(X\rightarrow p\bar{p})/B(X\rightarrow
\pi^+\pi^-\eta')\sim 1/3$. All of these support the interpretation
of that $X$-particle is a baryonium. In this talk I try to review
the works in Ref.\cite{Yan1}\cite{Yan2}, in which X(1835) has been
studied as proton-antiproton bound state in Skyrme model.

\section{Nucleon-Antinucleon Bound State in Skyrme Model}

\noindent Skyrme's old idea \cite{Skyr} that baryons are chiral
solitons has been successful in describing the static nucleon
properties \cite{Adkins} since Witten's illustration that the
soliton picture of baryons is consistent with QCD in the large $N_c$
approximation \cite{witt}. Actually, it is a part of the large
$N_c$-QCD theory. The Skyrme model has been widely used to discuss
baryons and baryonic-system properties. Deuteron is a typical
nucleon-nucleon system with baryon number 2, i.e., winding number 2,
which has been extensively studied in this model. Generally, there
are two Skyrmion ansatzes being used to explore this neutron-proton
bound state in the Skyrme model: the product ansatz proposed by A.
Jackson, A.D. Jackson and V. Pasquier\cite{Jackson}, and the
instanton ansatz proposed by M.F. Atiyah and N.S.
Manton\cite{Atiyah} for this non-trivial Skyrmion configurations.
However, for the deuteron-like system $p\bar{p}$, the winding number
(i.e., the baryon number in the Skyrme model)  is zero, therefore
the topology of the corresponding Skyrmion configurations is
trivial, and only the product ansatz works. In the follows, we
employ the ungroomed product ansatz to investigate the
$p\bar{p}$-system.

The Lagrangian for the SU(2) Skyrme model is
\begin{widetext}
\begin{equation}\label{su2}
    {\cal L}=\frac{1}{16}F^2_\pi {\mbox{Tr}}(\partial_\mu U^\dagger \partial^\mu
    U)+\frac{1}{32e^2}{\mbox{Tr}}([(\partial_\mu U) U^\dagger,(\partial_\nu
    U) U^\dagger]^2)+\frac{1}{8}m^2_\pi F^2_\pi {\mbox{Tr}}(U-1),
\end{equation}
\end{widetext}
 where  $U(t,\textbf{x})$ is the SU(2) chiral field, expressed in
 terms of the pion fields:
 \begin{equation}
    U(t,\mathbf{x})=\sigma(t,\mathbf{x})+i\pi(t,\mathbf{x})\cdot\mathbf{\tau}.
 \end{equation}
We fix the parameters $F_\pi$ and $e$ as in \cite{Adkins}, and our
units are related to conventional units via
\begin{equation}
\frac{F_{\pi}}{4e}=5.58~\mbox{MeV},     \ \
\frac{2}{eF_{\pi}}=0.755~\mbox{fm}.
\end{equation}
By using the hedgehog {\it Ansatz}
\begin{equation}
        U_{H}(\mathbf{r})=\exp[i\tau\cdot\widehat{r}f(r)],
\end{equation}
the ungroomed product {\it Ansatz} describing the behavior of
skyrmion and anti-skyrmion system without relatively rotation in the
iso-space is of the form
\begin{equation}
    U_s=U_{H}(\mathbf{r}+\frac{\rho}{2}\hat{z})
    U_{H}^\dagger(\mathbf{r}-\frac{\rho}{2}\hat{z}).
\end{equation}
 The skyrmion and
the anti-skyrmion are separated along the $\hat{z}$-axis by a
distance $\rho$.  We find that the static potential in the
ungroomed $S\bar S$ channel is physically more interesting, which
satiates
\begin{equation}
    U\rightarrow 1 \ \ \mbox{when} \ \ \rho\rightarrow 0.
\end{equation}
Fixing our {\it Ansatz} as above, we can get the static energy from
the skyrmion Lagrangian:
\begin{equation}\label{V}
    M(\rho)=\int d^3{\bf r}[
    -\frac{1}{2}{\mbox{Tr}}(R_{i}R_{i})-\frac{1}{16}{\mbox{Tr}}([R_{i},R_{j}]^2)],
\end{equation}
where $i,j=1,2,3$. The right-currents $R_{\mu}$ are defined via
\begin{equation}
    R_{\mu}=(\partial_{\mu}U)U^\dagger,
\end{equation}
and we express the energy in the units defined above. In this
picture, the binding energies for $S\bar{S}$ (which corresponds to
the classical binding energies of $p\bar{p}$) are
\begin{equation}\label{binding}
\Delta E_B=2m_p^c-M(\rho),
\end{equation}
where $m_p^c=867~\mbox{MeV}$ is the mass of a classical nucleon (or
classical Skyrmion). A stable or quasi-stable $p\bar{p}$-binding
state corresponds to the Skyrmion configuration
$U_s(r,\rho_B)=U_{H}(\mathbf{r}+\frac{\rho_B}{2}\hat{z})
    U_{H}^\dagger(\mathbf{r}-\frac{\rho_B}{2}\hat{z})$ with
$\Delta  E_B(\rho_B)<0$ and ${d \over d\rho}(\Delta
E_B(\rho))|_{\rho=\rho_B}=0.$

The numerical result of the static energy as a function of $\rho$ is
showed in Fig.~1. From it, we find that there is a quasi-stable
$p\bar{p}$-binding state:
\begin{eqnarray}\label{B1}
\rho_B &\approx& 2.5~\mbox{fm}, \\
\label{B2} \Delta E_B(\rho_B) &\approx& 10~\mbox{MeV}.
\end{eqnarray}
Obviously, this is a deuteron-like molecule state with rather small
binding energy, and hence its mass is rather close to the threshold
in proton-anti-proton mass spectrum. In this case, as long as one
could further prove that this quasi-stable molecule state could have
indeed a decay width, say $\sim 10~\mbox{MeV}$, then one could
naturally conclude that this molecule resonance did respond for the
near-threshold enhancement in $(p\bar{p})$-mass spectrum from
$J/\psi\rightarrow \gamma p \bar{p}$ observed by BES, and this
deuteron-like  state should be the  the particle reported in
experiment of ref.\cite{BES}. We shall do so in the next section.

\begin{figure}[hptb]
   \centerline{\psfig{figure=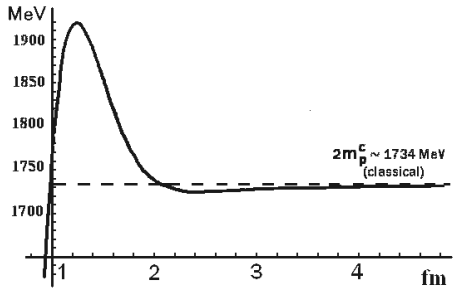,width=8cm}}
 \centering
 \vskip-0.5in
\begin{minipage}{3.45in}
    \caption{The static energy of skyrmion-antiskyrmion
    system, where $m_p^c$ is the classical single skyrmion mass without quantum correction.}
\end{minipage}
\end{figure}
\noindent {Three} remarks are in order:

1) To the $p\bar{p}$-bound state, { the absolute baryon number for
separate $(N-\bar{N})$ system is useful for understanding the inside
structure of $X$-particle in the Skyrmion framework. } By the Skyrme
model, for $(N-\bar{N})$-system, $U(\mathbf{r},\rho)\equiv
U_{H}(\mathbf{r}+\frac{\rho}{2}\hat{z})
U_{H}^\dagger(\mathbf{r}-\frac{\rho}{2}\hat{z})$. The baryon density
reads
\begin{eqnarray} \nonumber \rho_B(\mathbf{r},
\rho)&=&{\epsilon^{ijk}\over 24\pi^2} Tr[(U^\dag(\mathbf{r},
\rho)\partial_iU(\mathbf{r}, \rho))(U^\dag(\mathbf{r},
\rho)\partial_jU(\mathbf{r}, \rho))  \\
\label{density}&\times &(U^\dag(\mathbf{r},
\rho)\partial_kU(\mathbf{r}, \rho))],
\end{eqnarray}
and  the total baryon number of the system with any separation
$\rho$ is zero, i.e.,
\begin{equation}\label{B}
 B(\rho)\equiv \int
d^3\mathbf{r}\rho_B(\mathbf{r}, \rho)=0.
\end{equation}
But obviously the absolute baryon number for separate $(N-\bar{N})$
is nonzero,
\begin{equation}\label{|B|}
{ |B(\rho)|\equiv \int d^3\mathbf{r}|\rho_B(\mathbf{r}, \rho)|\neq
0}.
\end{equation}
 From the bottom panel of Fig.2, one can see when $N$ and $\bar{N}$
 are separated away largely, saying $\rho>  1.5fm$,
$|B(\rho)|\simeq 2$, and it is almost $\rho$-independent. When
$\rho< 1.5fm$, $|B(\rho)|$ decreases sharply, and when $\rho< 0.5fm$
the "baryonic matters" in the system are almost all annihilated
away, i.e., $|B(\rho)|\rightarrow 0$. The curve of function
$|B(\rho)|$ indicates that the configurations of $N\bar{N}$-Skyrmion
with $\rho_{N\bar{N}}>1.5fm$ are molecular  states of $N-\bar{N}$,
and ones with $\rho_{N\bar{N}}<0.5fm$, they are mesonic states. The
upper panel of Fig.2 shows the curve of the static energy of
Skyrmion-baryoniums (the same as Fig.1). Obviously, since the
distance  between $p$ and $\bar{p}$ inside $X$ $\rho_B\approx 2.5fm$
(see Eq.(\ref{B1})) is much larger than $\sim 1.5fm$, and
$|B(\rho_B)|\simeq 2$, we conclude this baryonium configuration
corresponding to the enhancement belongs to a molecular bound state
consisted of proton and antiproton.

\begin{figure}[hptb]
   \centerline{\psfig{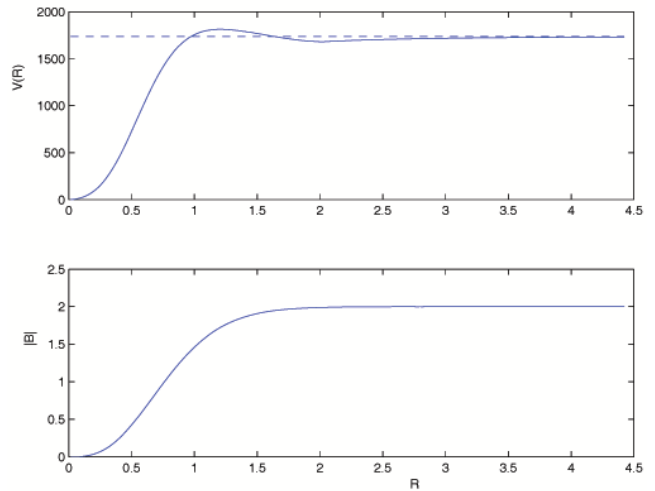}}
 \centering
 \vskip-0.5in
\begin{minipage}{3.45in}
\vskip0.5in
    \caption{The static energy of skyrmion-antiskyrmion
    system  and the absolute value of baryon number of $N\overline{N}$:
    1, The upper panel shows the curve of the static energy of skyrmion-antiskyrmion
    system (the same as Fig.1); 2, The bottom panel shows the absolute value of baryon
    number of $N\overline{N}$: $|B(\rho)|$. }
\end{minipage}
\end{figure}

2) In above, the discussion on the binding energies is somehow
qualitative. Only the classical soliton energies of $p\bar{p}$ have
been taken into account there. Even though they should be the
leading order of large $N_c$ expansion, the semi-classical quantum
correction to the $p\bar{p}$ energies may decrease the  small
binding energies so as to vanish. Similar problem occurred in the
discussions to deuteron by using $\tau_3$-groomed product skyrmion
ansatz\cite{Jackson}. There were two methods to get over this
difficulty for deuteron case: changing the ansatz\cite{Atiyah}, or
improving the calculations of the classical soliton energies to
include the contributions from the higher order of $N_c$-expansion
(or higher order of derivative expansion)\cite{Mau}. The  Skyrme
model with 6-order derivative term has been explored in order to
calculate the the quark spin contents in proton\cite{LY1,LY2}, and
hence the model has been fixed. It should be interesting to pursue
the the semi-classical quantum corrections
 to the $p\bar{p}$ in this generalized Skyme model. Namely, one could
learn  how to quantize the classical baryonium solutions without
topology charge (see Eq.(\ref{B})) semi-classically.

3) Eq.(\ref{su2}) is the $SU(2)$-Skyrmion Lagrangian. It can be
extended  into $SU(3)$-one when $U(x)$ becomes into $3\times
3$-matrix function, the Wess-Zumino term is added, and the the
$SU(2)$ chiral symmetry breaking term in (\ref{su2}) changes to
$SU(3)$-one\cite{su3,su31}. The real world has three light flavors,
and hence there are {\it three} flavor contents at least in the
"sea" of $SU(2)$-nucleons\cite{su31}. The hyper-baryons with strange
quantum number, of course, only emerge in the $SU(3)$-Skyrmion
theory. The fact, however, that $SU(2)$-Skyrmion can rather rightly
describe the properties of real nucleons indicates the three flavor
"sea" effects of nucleons have been partly (at least) covered by
$SU(2)$-Skyrmion. The $(p-\bar{p})$-system discussed in this paper
is flavor singlet and the baryon number free, and hence the
qualitative discussions on it by means of $SU(2)$-Skyrmion should
consistent with one by $SU(3)$-Skyrmion.

\section{a phenomenological model with a skyrmion-type
potential}

\noindent In order to further catch the features of the Skyrmion
prediction to $p\bar{p}$ and to derive the decay width of this
quasi-stable particle, we employ the potential model  induced from
the skyrmion picture of $p\bar{p}$-interactions for the nucleons.
For over fifty years there has been a general understanding of the
nucleon-nucleon interaction as one in which there is, in potential
model terms, a strong repulsive short distance core together with a
longer range weaker attraction. The attractive potential at the
middle range
 binds the neutron and the proton to form a deuteron.
 In comparison with the skyrmion result on the deuteron
\cite{b_c,Schramm,a_m,lms}, we notice several remarkable features
of the static energy $M( \rho)$ and the corresponding
$(p\bar{p})$-potential $V(\rho)$ . Firstly, the potential is
attractive at $\rho>2.0~\mbox{fm}$, similar to the deuteron case.
This is due to the reason that the interaction via pseudoscalar
$\pi$-meson exchange is attractive for both quark-quark $qq$ and
quark-antiquark $q\bar{q}$ pairs. Physically, the attractive force
between $p$ and $\bar{p}$ should be stronger than that of $p n$.
Therefore the fact that our result of the $p\bar{p}$-binding
energy (see eq.(\ref{B2})) $\Delta E_B(\rho_B) (\approx
10~\mbox{MeV})$ is larger than that of deuteron
($2.225~\mbox{MeV}$) is quite reasonable physically. Secondly,
there is a static Skyrmion energy peak at $\rho\sim 1~\mbox{fm}$
in Fig.~1. This means that the corresponding potential between $p$
and $\bar{p}$ is repulsive at that range. This is an unusual and
also an essential feature. The possible explanation for it is that
the skyrmions are extended objects, and there would emerge a
repulsive force to counteract the deformations of their
configuration shapes when they close to each other.  Similar
repulsive potential has also been found in previous numerical
calculation \cite{lu_am}. Thirdly, a well potential at middle
$\rho$-range is formed due to the competition between the
repulsive and attractive potentials mentioned above, similar to
the deuteron case. But the depth should be deeper than that in the
deuteron case, as argued in a QCD based discussion
\cite{Datta}\cite{Yan3}. The $p$ and $\bar{p}$ will be bound to
form a baryonium in this well potential. Finally, the potential
turns to decrease quickly from $\sim 2000~\mbox{MeV}$ to zero when
$\rho \to 0$. This means that there is a strong attractive force
at $\rho \sim 0$.
Physically, the skyrmions are destroyed at this $\rho$ range, and
$p\bar{p}$ are annihilated.

\begin{figure}[hptb]
\centerline{\psfig{figure=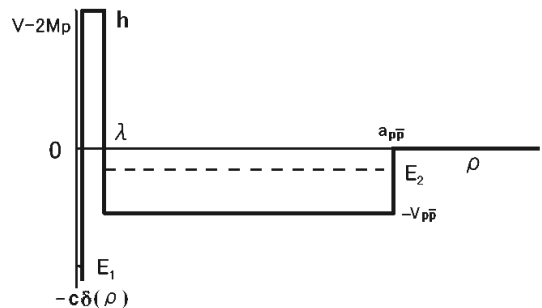,width=7.5cm}}
 \centering
\vskip-0.3in
\begin{minipage}{3.45in}
    \caption{The skyrmion-type potential of $p\bar{p}$-system.}
\end{minipage}
\end{figure}

The qualitative features of the proton-neutron potential for the
deuteron can be well described by a simple phenomenological model
of a square well potential 
\cite{Ma,Hulthen,Schiff} with a depth which is sufficient to bind
the $pn$ ${^3S_1}$-state with a binding energy of
$-2.225~\mbox{MeV}$. Numerically, the potential width $a_{pn}$ is
about $2.0~\mbox{fm}$, and the depth is about
$V_{pn}=36.5~\mbox{MeV}$. Similarly, from the above illustrations on
the features of the potential between  $p$ and $\bar{p}$ based on
the Skyrmion picture, we now construct a phenomenological potential
model for the $p\bar{p}$ system, as shown in Fig.~3, and it will be
called as skyrmion-type potential hereafter.

We take the width of the square well potential, denoted as
$a_{p\bar{p}}$, as close to that of the deuteron,  i.e.,
$a_{p\bar{p}}\sim a_{pn}\simeq 2.0 ~\mbox{fm}$. According to QCD
inspired considerations \cite{Datta,Yan3,Maltman,Rujula}, the well
potential between $q$ and $\bar{q}$ should be double (or more)
attractive than the $qq$-case, i.e., the depth of the $p\bar{p}$
square well potential is $V_{p\overline{p}}\simeq
2V_{pn}=73~\mbox{MeV}$. The width for the repulsive force revealed
by the Skyrme model can be fitted by the decay width of the
baryonium, and we take it to be $\lambda = 1/(2m_p) \sim
0.1~\mbox{fm}$, the Compton wave length of the bound state of
$p\bar{p}$. The square barrier potential begins from $\rho \sim
\lambda$, and the height of the potential barrier, which should be
constrained by both the decay width and the binding energy of the
baryonium, is taken as $2m_p+h$, where $h \sim m_p/4$. At $\rho\sim
0$, $V^{(p\bar{p})}(\rho)\sim -c\;\delta (\rho)$ with a constant
$c>0$.

Analytically, the potential $V(\rho)$ is expressed as follows
\begin{equation}\label{potential}
V(\rho)=2m_p-c\;\delta(\rho)+V_c(\rho),
\end{equation}
where
\begin{equation}
V_c(\rho)=\left\{
   \begin{array}{ll}
   {h={m_p/4}},\ \ & 0<\rho<\lambda,\\
   -V_{p\bar{p}}
   =-73~\mbox{MeV},     & \lambda<\rho<a_{p\bar{p}},\\
   0, &~~~~~\rho>a_{p\bar{p}}.
   \end{array} \right.
\end{equation}
With this potential, the Schr$\ddot{o}$dinger equation for $S$-wave
bound states is
\begin{equation}\label{schrodinger}
{-1\over 2(m_p/2)}{\pa^2\over \pa \rho^2}
u(\rho)+\left[V(\rho)-E\right]u(\rho)=0,
\end{equation}
where $u(\rho)=\rho\;\psi(\rho)$ is the radial wave function,
$m_p/2$ is the reduced mass. This equation can be solved
analytically, and there are two bound states $u_1(\rho)$ and
$u_2(\rho)$ (see Fig. 4 and Fig. 5): $u_1(\rho)$ with binding energy
$E_1< -V_{p\bar{p}}=-73~\mbox{MeV}$ is due to
$-c\;\delta(\rho)$-function potential mainly, and $u_2(\rho)$ with
binding energy $E_2>-73~\mbox{MeV}$ is due to the attractive square
well potential at middle range mainly. $u_1(\rho)$ is the vacuum
state. And, clearly, $u_2(\rho)$ should correspond to a
deuteron-like molecule state and it may be interpreted as the new
$p\bar{p}$ resonance reported by BES~\cite{BES}: $X(1835)$. It is
also expected that corresponding binding energies $-E_2$ in the
potential model provided in above $\Delta E_B(\rho_B)$ (see
eq.(\ref{B2})) are all in agreement of the data within errors of
BES~\cite{BES}. By fitting experimental data, we have
\begin{eqnarray}\label{E1} E_1&=&-(2m_p-m_{\eta_0})\simeq
-976~\mbox{MeV},\\ \label{E2} E_2&=&-17.2~\mbox{MeV}.
\end{eqnarray}
Considering its decay width will be derived soon (see
eq.(\ref{GG})), we conclude that the near-threshold narrow
enhancement in the $p\bar{p}$ invariant mass spectrum from
$J/\psi\rightarrow\gamma p
\bar{p}$ 
might be interpreted as a state of protonium in this potential
model.

\begin{figure}[hptb]
\begin{center}
\includegraphics*[5pt,5pt][200pt,165pt]{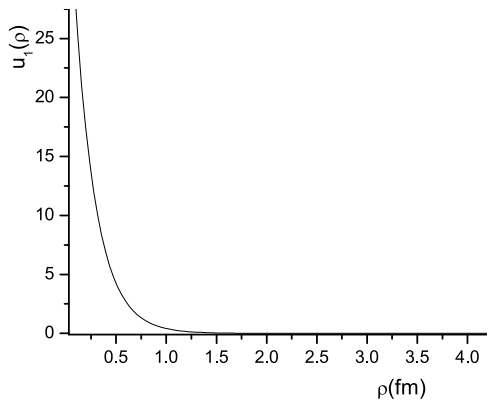}
\vskip-0.3in \caption{The Wave Function $u_1(\rho)$.}
\end{center}
\end{figure}
\vskip-0.3in
\begin{figure}[hptb]
\begin{center}
\includegraphics*[5pt,5pt][200pt,205pt]{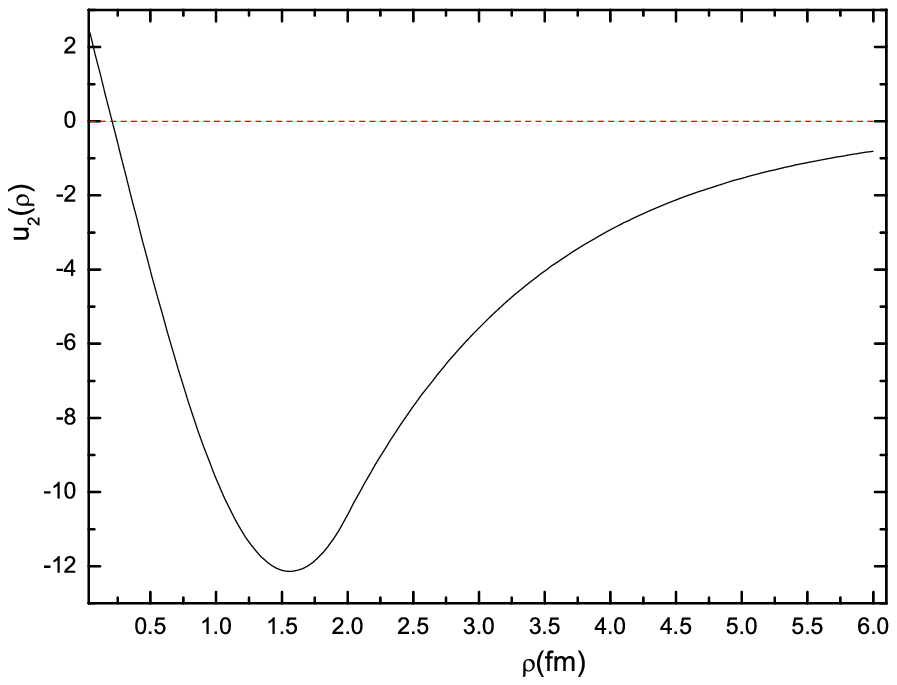}
\vskip-0.3in\caption{The Wave Function $u_2(\rho)$.}
\end{center}
\end{figure}

In the skyrmion-type potential of $p\bar{p}$, there are two
attractive potential wells: one is at $\rho\sim 0$ and the other is
at middle scale, together with a potential barrier between them. At
$\rho\sim 0$, the baryon- and anti-baryon annihilates. Naturally, we
postulate that the bound states decay dominantly by annihilation
and, therefore, we can derive the width of protonium state
$u_2(\rho)$ by calculating the quantum tunnelling effect for
$u_2(\rho)$ passing through the potential barrier. By
WKB-approximation, the tunnelling coefficient (i.e., barrier
penetrability) reads \cite{Schiff}
\begin{eqnarray}\label{WKB}
T_0&=&\exp\left[-2\int_0^{\lambda} dr \sqrt{m_p(h-E_2)}\right]
\nonumber \\  &=&\exp \left[-2\lambda\sqrt{m_p(h-E_2)}\right].
\end{eqnarray}
In the square well potential from $\lambda$ to $a_{p\bar{p}}$, the
time-period $\theta$ of round trip for the particle is
\begin{equation}
\label{trip} \theta={2\left[a_{p\bar{p}}-\lambda\right]\over
v}=\left[a_{p\bar{p}}-\lambda\right]\sqrt{{m_p\over
V_{p\bar{p}}+E_2}}.~
\end{equation}
Thus, the state $u_2(r)$'s life-span is $\tau=\theta T_0^{-1}$, and
hence the width of that state reads
\begin{eqnarray}
\label{Gamma} \Gamma\equiv {1\over \tau}
&=& {1\over a_{p\bar{p}}-\lambda} \sqrt{{ V_{p\bar{p}}+E_2 \over
m_p}}
\nonumber \\
&~ & \exp \left[-2\lambda\sqrt{m_p(h-E_2)}\right].~~
\end{eqnarray}
Numerically, substituting
$E_2=-17.2~\mbox{MeV},\;a_{p\bar{p}}=2.0~\mbox{fm}$ into
(\ref{Gamma}), we obtain the prediction of $\Gamma$:
\begin{equation}\label{GG}
\Gamma\simeq 15.5~\mbox{MeV},
\end{equation}
which is a reasonable number when comparing it with the experimental
data \cite{BES}. This result  indicates also that the
$(p\bar{p})$-collision times per second in $X$ are about
\begin{equation}\label{nu}
\nu=\Gamma/\hbar\simeq 2.35\times 10^{22}~~{\rm times/second}.
\end{equation}
This reflects the possibility of $(p\bar{p})$-collisions inside
$X$-particle in the sense of quantum theory. When that possibility
$P=1=\nu\tau$, one reobtains the lifetime of $X$ as
$\tau=1/\nu=1/\Gamma$. Noting since the binding energy $E_2$ is
rather small (comparing with $2m_p$), the annihilations which cause
$X(1835)$ to be unstable are almost at rest.

Two remarks are in order:

1)  Because there are adjustable parameters $(
c,\;h,\;\lambda,\;V_{p\bar{p}},\;a_{p\bar{p}})$ in our potential
model, it is no doubt to fit the renewed experimented data in
Ref.\cite{BES2}.

 2, The  quantum numbers of
$X$-particle corresponding to $u_2(\rho)$ are
$I^G(J^{PC})=0^+(0^{-+})$\cite{Datta,GZh}, and hence the particle
corresponding to $u_1(\rho)$ must be of $I^G(J^{PC})=0^+(0^{-+})$
also. Considering the range for nonzero $u_1(\rho)$ is  of $\rho
\sim 0$ (see Fig.4), and then $|B(\rho \sim 0)|\sim 0$ (see Fig. 2),
the $u_1(\rho)$-particle must be a meson (a point-like particle)
rather than a baryonim. Thus, the possible candidates are $\eta$ or
$\eta'$. Moreover, the gluon contents for $u_1(\rho)$-particle,
$u_2(\rho)$-particle $X(1835)$ and $(p\bar{p})$ should be same.
Thus, only $\eta'$ is available, and then we  conjectured in above
that the particle corresponding to $u_1(\rho)$ is $\eta'$.

\section{Proton-Antiproton Annihilation inside X(1835) and its
mesonic decays}

\noindent In above, we have pointed out that $N\bar{N}$-annihilation
causes the $p\bar{p}$ baryonium $X(1835)$ to be unstable. In order
to find out what decay processes are of the most favorable channels,
we now pursue the $N\bar{N}$ annihilations inside  $X(1835)$ in the
Skyrme model. As a nucleon model inspired by QCD, the Skyrme model
has a very useful advantage in describing $N\bar{N}$ annihilation:
this effective theory provides a self-contained dynamics that
encompasses nonlinear processes such as meson production, baryon
excitation, and annihilation. The Skyrme model requires no
additional dynamical assumption, such as the {\it ad hoc} dynamical
behavior of the color-confinement wall that must be assumed in bag
models. The studies of annihilation in the Skyrme model\cite{somm,
shao} have suggested that annihilation proceeds very rapidly when
the baryon and antibaryon collide and that the product of this rapid
annihilation is a pion pulse or coherent pion wave. The annihilation
of $N\bar{N}$ scattering process at rest has been investigated by
Amado, Cannata, Dedoder, Locher, Shao and Lu
(ACDLSL)\cite{batch1,Yang lu3,batch2} in the Skyme model by using
the coherent state method. When the proton and antiproton collided,
they will be annihilated into mesons rapidly.
$(p\bar{p})$-annihilation at rest but without considering the $P-$
and $G-$parity has been investigated by using coherent state method
in \cite{batch1, Yang lu3,batch2}. In the follows, we introduce
ACDLSL's coherent state of Refs.\cite{batch1, Yang lu3,batch2}
briefly, and then, the studies in Ref.\cite{Yan2} which lead to
reveal the most favorable channels for $X$-decays.

\subsection{coherent state for annihilation of $N\bar{N}$ scattering
process at rest.} \noindent 1) Coherent state is the eigenstate of
annihilation operator, $a|\lambda\rangle =\lambda |\lambda\rangle$:
\begin{equation}\label{A1} |\lambda\rangle=e^{-|\lambda|^2/2}e^{\lambda
a^\dag}|0\rangle,\end{equation} which is useful for the quantum
optics, e.g., to describe the purse from rapid radiations of photons
in laser\cite{Glauber}.
 {To free quantum
scalar field} $ \phi(x)=\phi^{(+)}(x)+\phi^{(-)}(x)= \int
{d^3\mathbf{k} \over (2\pi)^{3/2}\sqrt{2\omega_{\bf k}}} (a_{\bf
k}e^{-ik\cdot x}+ a^{\dag}_{\bf k}e^{ik\cdot x}), $ the normalized
quantum  state $|f\rangle $ defined by
\begin{equation}\label{A2}|f\rangle=\exp \left[-{1\over 2} \int
d^3\mathbf{k} |f({\bf{k}})|^2 +\int d^3{\bf{k}} f({\bf{k}})
a^{\dag}_{\bf k}\right] |0\rangle, \end{equation} which is { the
coherent state of $\phi(x)$}, i.e., \begin{eqnarray}\nonumber
\phi^{(+)}(x)|f\rangle &=& \varphi(x)|f\rangle, ~~~with\\ \label{A3}
\varphi(x)&=&\int \frac{d^3\mathbf{k}} {
(2\pi)^{3/2}\sqrt{2\omega_{\mathbf{k}}} } e^{-ik\cdot x} f(\bf{k}).
\end{eqnarray}

\noindent 2) { The coherent state with fixed 4-momentum and
isospin\cite{batch1,Yang lu3,batch2}:} The Fig.2 shows when the
distance  between Skyrmion and anti-Skyrmion $\rho$ less than $\sim
1fm$, $|B(\rho)|$ will decrease sharply. This means the pions
radiated from the annihilation of $N\bar{N}$ form a purse. The
processes are very rapid, and are similar to the photon radiated
from the laser. Considering this feature of $N\bar{N}$-annihilation,
a coherent state description to the $N\bar{N}$ scattering
annihilation has been suggested in Ref.\cite{batch1}, and a coherent
state with fixed 4-momentum and isospin\cite{batch1,Yang lu3,batch2}
has been formed by Amado, Cannata, Dedoder, Locher, Shao and Lu
(ACDLSL) as follows

\begin{eqnarray}\label{ACD}
\label{1.35}|K,I,I_z\rangle_{\rm ACDLSL}=\hskip-0.05in\int
\hskip-0.05in \frac{d^4x}{(2\pi)^4}
\frac{d\Omega_{\hat{\textbf{n}}}}{\sqrt{4\pi}}~e^{iK\cdot
x}|f,x,\hat{\textbf{n}},2\rangle Y^{*}_{I,I_z}(\hat{\textbf{n}})
\end{eqnarray}
where
\begin{eqnarray}\label{ACD1}
|f,x,\hat{\textbf{n}},2\rangle&=&[e^{F(x,\hat{\textbf{n}})}
-F(x,\hat{\textbf{n}})-1 ]|0\rangle, \\
\nonumber F(x,\hat{\textbf{n}})&=&\int d^3{\bf k} e^{-ik\cdot x}
f({\bf k})a_{\bf k}^\dag \cdot \hat{\bf n},\\
\label{ACD2} |f(\textbf{k})|^2&=&\frac{C~
\textbf{k}^2}{(\textbf{k}^2+\alpha
^2)^2(\omega^2_{\textbf{k}}+\gamma^2)^2 \omega^2_{\textbf{k}}}
\end{eqnarray}
where $\alpha=\gamma=2m_\pi.$ Then the $\pi$-radiations from
$|K,I,I_z\rangle_{\rm ACDLSL}$ can be discussed by calculating the
mean number of charged $\pi^\pm$ and $\pi^0$ ($\mu=\pm,\;0$):
\begin{eqnarray*}
\mathbf{\hat{N}}_{\mu}&=&_{\rm ACDLSL}\langle K,I,I_{z}|\int d^{3}k
a^{\dag}_{\mathbf{k},\mu}a_{\mathbf{k},\mu}|K,I,I_{z}\rangle_{\rm
ACDLSL}.
\end{eqnarray*}
The function $|f(\textbf{k})|^2$ in eq.(\ref{ACD2}) comes from the
following considerations provided in Ref.\cite{Yang lu3}. The
$\pi$-field equation is:
\begin{equation}\label{ACD3}
\left( \nabla^2-{\pa^2 \over \pa t^2}-\mu^2 \right)\Phi({\bf r},t)
=S({\bf r},t).
\end{equation}
Here $S({\bf r}, t)$ is the source of the pion field  $\Phi$.
Inspired by the Skyrmion calculations\cite{somm, shao}, a very
simple spherically symmetric form of $S({\bf r},t)$ has been
suggested in Ref.\cite{batch1, Yang lu3} by Amado, Cannata,
Dedonder, Locher and Shao (i.e., ACDLS-{\it ansatz}) as follows
\begin{equation}\label{ACD4}
S({\bf r},t)=S({\bf r},t)_{ACDLS}=\left\{\begin{array}{cc}
0, & {\rm if}~ t<0, \\
S_0{e^{-\alpha r} \over r}te^{-\gamma t}, & {\rm if}~ t>0.
\end{array} \right.
\end{equation}
 By using $f({\bf k})=-i\sqrt{2\pi}S({\bf
k}, \omega_k)/(2\omega_k)$ and
$$ S({\bf k}, \omega_k)=\int {d^3{\bf r}\; dt \over (2\pi)^2}\exp (-i{\bf
k\cdot r}+i\omega_k t) S({\bf r},t),$$ we get the eq.(\ref{ACD2}).

\subsection{Coherent states with fixed $G-$ and $P-$parities}
\noindent We address that  the coherent state
$|K,I,I_{z}\rangle_{\rm ACDLSL}$ is $G-$ and $P-$parities free,
therefore it can not be used to discuss the $X(1835)$-decays. In
Ref.\cite{Yan2}, $X(1835)$ has been treated as meson radiation
source with $I^{G}(J^{PC})=0^{+}(0^{-+})$.  The coherent state with
fixed 4-momentum, isospin,  G-parity($+$) and P-parity($-$) has been
constructed in Ref.\cite{Yan2} as follows
\begin{eqnarray}\label{GP1}
|K,I,I_z\rangle_{GP}=\hskip-0.05in\int \hskip-0.05in
\frac{d^4x}{(2\pi)^4}
\frac{d\Omega_{\hat{\textbf{n}}}}{\sqrt{4\pi}}~e^{iK\cdot
x}|f,g,x,\hat{\textbf{n}},2\rangle Y^{*}_{I,I_z}(\hat{\textbf{n}})
\end{eqnarray}
where
\begin{eqnarray*}\label{GP2}
|f,g,x,\hat{\textbf{n}},2\rangle
 &\hskip-0.1in=&\hskip-0.1in[e^{F(x,\hat{\textbf{n}})}
+e^{G(x,\hat{\textbf{n}})}-F(x,\hat{\textbf{n}})-G(x,\hat{\textbf{n}})\\
-e^{-F(x^{'},\hat{\textbf{n}})}&&\hskip-0.2in-e^{-G(x^{'},\hat{\textbf{n}})}
-F(x^{'},\hat{\textbf{n}})-G(x^{'},\hat{\textbf{n}})]|0\rangle,\\
 F(x,\hat{\textbf{n}})=\int d^3 {\bf k} &&\hskip-0.2in e^{-ik\cdot
x}f(\textbf{k})\textbf{a}^{\dag}_{\textbf{k}}\cdot\hat{\textbf{n}}
+\int d^3 {\bf q} e^{-iq\cdot x} g(\textbf{q})
b^{\dag}_{\textbf{q}},\\
G(x,\hat{\textbf{n}})=-\int  d^3 {\bf k} &&\hskip-0.2in
f(\textbf{k})
\textbf{a}^{\dag}_{\textbf{k}}\cdot\hat{\textbf{n}}~~e^{-ik\cdot
x}+\int d^3 {\bf q} e^{-iq\cdot x} g(\textbf{q})
b^{\dag}_{\textbf{q}},
\end{eqnarray*} \vskip-0.3in
\begin{eqnarray}\label{GP01}
 |f(\textbf{k})|^2&=&  \frac{C~
\textbf{k}^2}{(\textbf{k}^2+\alpha
^2)^2(\omega^2_{\textbf{k}}+\gamma^2)^2 \omega^2_{\textbf{k}}}, \\
\label{GP02}|g(\textbf{k})|^2
&=&|f(\textbf{k})|^2|_{m_\pi\rightarrow m_\eta}
\end{eqnarray}
where $\alpha=\gamma=2m_\pi,$ and $x'=(-{\bf x}, t)$. Because both
$\pi$- and $\eta$-fields are pseudo-Goldstone bosons in QCD, the
ACDLS-{\it ansatz}  eq.(\ref{ACD4}) has been used for both $\pi$-
and $\eta$-particle radiations to determine $|f(\textbf{k})|^2$ and
$|g(\textbf{k})|^2$ in Eqs.({\ref{GP01}) and (\ref{GP02}). $C$ is
the strength and can be fixed by requiring that the average energy
is the energy released in annihilation, which is equal to $2m_{p}$,
i.e.,
\begin{equation}
2m_p=\int d^3{\bf k}\left(\sqrt{{\bf k}^2+m_\pi^2} |f({\bf k})|^2
+\sqrt{{\bf k}^2+m_\pi^2} |g({\bf k})|^2 \right),
\end{equation}
where the strengthes of $C$ in both $|f({\bf k})|$ and $|f({\bf
k})|$ are same because  both $\pi$ and $\eta$ are pseudo-Goldstone
bosons in QCD, and the influence to $C$ duo to light quark flavor
symmetry breaking is ignored (noting the effects of the light quark
flavor symmetry breaking to $|f({\bf k})|$ and $|f({\bf k})|$ are
taken into account via $m_\pi\neq m_\eta$).
 The unitary operators $\hat{G}$,~$\hat{P}$  can be expressed
as follows
\begin{eqnarray*}
\nonumber
\hat{P}&=& \exp[i\frac{\pi}{2}\sum_{j,\textbf{k}}(a^{+}_{\textbf{k},j}a_{-\textbf{k},j}+b^{+}_{\textbf{k}}b_{-\textbf{k}}+a^{+}_{\textbf{k},j}a_{\textbf{k},j}+b^{+}_{\textbf{k}}b_{\textbf{k}})]\\
&& \hskip-0.4in \hat{G} = \exp[i\frac{\pi}{2}\sum_{j,\textbf{k}}
(a^{+}_{\textbf{k,-1}}a_{\textbf{k},1}+a^{+}_{\textbf{k,1}}a_{\textbf{k},-1}
-a^{+}_{\textbf{k},1}a_{\textbf{k,1}}-a^{+}_{\textbf{k},-1}a_{\textbf{k,-1}})]\\
&&\hskip-0.3in \times
\exp[-\frac{\pi}{\sqrt{2}}\sum_{\textbf{k}}(a^{+}_{\textbf{k},0}a_{\textbf{k,1}}
+a^{+}_{\textbf{k},0}a_{\textbf{k,-1}}-a^{+}_{\textbf{k},1}a_{\textbf{k},0}
-a^{+}_{\textbf{k},-1}a_{\textbf{k},0})].
\end{eqnarray*}
It is easy to check the following equations
\begin{eqnarray}
\nonumber \hat{G}a^{\dag}_{\textbf{p},i}\hat{G}^{\dag}&=&-a^{\dag}_{\textbf{p},i }\\
\nonumber \hat{G} b^{\dag}_{\textbf{q}}\hat{G}^{\dag}&=&b^{\dag}_{\textbf{q}}\\
\nonumber \hat{P} a^{\dag}_{\textbf{p},i}\hat{P}^{\dag}&=&-a^{\dag}_{-\textbf{p},i }\\
\label{1.32} \hat{G}
b^{\dag}_{\textbf{q}}\hat{G}^{\dag}&=&-b^{\dag}_{-\textbf{q}}
\end{eqnarray}
where $i=1,0,-1$ corresponding to $\pi^{+},\pi^{0},\pi^{-}$, and
$b^\dag_q$ represents the creation operator for a pseudo-scalar
Goldstone particle with $G$-parity $(+)$, $P$-parity $(-)$ and
isospin $I=0$. So, it is $\eta$.

\subsection{$\pi$- and $\eta$-radiations from $| K, I,
I_z\rangle_{GP}$:} \noindent   The probability of ($N_{\pi}\pi$,
$N_{\eta}\eta$)-radiations from $| K, I, I_z\rangle_{GP}$ is:
\begin{eqnarray}
\nonumber
P(N_{\pi},N_{\eta})&=&\frac{1}{N_{\pi}!N_{\eta}!}\int\prod_{i=1}^{N_{\pi}}d^3{\bf
p}_i \prod_{j=1}^{N_{\eta}}d^3{\bf q}_j\\ \nonumber && |\langle
\textbf{p}_1 \textbf{p}_2 \cdot\cdot\cdot \textbf{p}_{N_{\pi}}
\textbf{q}_1 \textbf{q}_2\cdot\cdot\cdot
\textbf{q}_{N_{\eta}}|K,I,I_z\rangle_{GP}|^2\\
 \label{1.50} &=&\frac{1}{{\cal I}(K)}\frac{16
I(K,N_{\pi},N_{\eta})F(N_{\pi},I)}{N_{\pi}!N_{\eta}!}.
\end{eqnarray}
where ${\cal I} (K)$ is the normalization factor:
$$
\label{1.40}{\cal I}(K)=\sum_{m+n \geq 2;\;\; m {\rm \;is\; even},\;
n {\rm \;is\; odd}} \frac{16~I(K,m,n)}{m! n!}F(m,I)
$$
where
\begin{eqnarray*}
\label{1.41}I(K,m,n)&=& \int
\delta^{4}(K-\sum_{i=1}^{m}p_i-\sum_{j=1}^{n}q_j)
\prod_{i=1}^{m}d^3\textbf{p}_i|f(\textbf{p}_i)|^2 \\
&\times &\prod_{j=1}^{n}d^3\textbf{q}_j|g(\textbf{q}_j)|^2
\end{eqnarray*}
and
\begin{eqnarray*}
\nonumber F(m,I)&=&\int
\frac{d\Omega_{\hat{\textbf{n}}}d\Omega_{\hat{\textbf{n}}^{'}}
}{4\pi}Y^{*}_{I
I_z}(\hat{\textbf{n}})Y_{II_z}(\hat{\textbf{n}}^{'})(\hat{\textbf{n}}\cdot\hat{\textbf{n}}^{'*})^{m}\\
\label{1.42}& =&\left\{   \begin{array}{ll}
0 & I >  m \mbox{ or } I-m \mbox{ is odd} \\
\frac{ m! } { (m-I)!! (I+m+1)!! } & I\le m \mbox{ and } I-m \mbox{
is even}.
\end{array}   \right.
\end{eqnarray*}

 { Noting the branching fraction
$B(X(1835)\rightarrow m\pi+n\eta)\propto P(m,n)$, the ratios  of
$B(X\rightarrow \eta 4\pi)/B(X\rightarrow \eta 2\pi),~~etc$ can be
calculated:}
\begin{eqnarray}\label{result1}
\frac{B(X\rightarrow \eta4\pi)}{B(X\rightarrow \eta
2\pi)}&=&\frac{I(K,4,1)}{I(K,2,1)}\frac{7}{300}\simeq 1.8\times
10^4,\\
 \label{result2} \frac{B(X\rightarrow \eta2\pi)}{B(X\rightarrow 3\eta
)}&=&\frac{I(K,2,1)}{I(K,0,3)}\frac{2}{3}\simeq 5.9.
\end{eqnarray}
These are desirous results. The process of $X\rightarrow \eta 4\pi$
is a very favorable channel, and the branching fraction of the
simplest decay channel $X\rightarrow \eta 2\pi$ is suppressed by
about 4 orders comparing with one of $X\rightarrow \eta 4\pi$.

 \subsection{ An intuitive picture:}
\noindent 1)Naively, the number of valence quarks in $X(1835)$(or
$(p\bar{ p})$) is equal to  the number of valence quarks in
$(\eta\pi\pi).$

\noindent 2)However, the gluon content for $(p\bar{p})$ and
$(\eta\pi\pi)$ are different.  Skyrmion tell us  the gluon
mass-percentage in the proton (or antiproton) is larger than that
for the $\pi$ or $\eta$.

\noindent 3)This indicates that there are some "redundant gluons"
that are left over in the process of $X\equiv(p\bar{p})\rightarrow
\eta 2\pi$. Consequently, the process should be expressed as
\begin{eqnarray*}\label{end1}
X\equiv(p\bar{p})\rightarrow \eta2\pi G.
\end{eqnarray*}
 where $G$ represents the "redundant gluons".
Going further
\begin{eqnarray*}\label{end1}
 X\equiv(p\bar{p})\rightarrow \eta2\pi G
\end{eqnarray*}
\vskip-0.45in
$$\hskip0.85in |$$ \vskip-0.53in \begin{equation}\label{end} \hskip1.27in \longrightarrow
\pi\pi \end{equation} Therefore, the most favorable $X$-particle
decay process should be $X\rightarrow \eta 4\pi$, instead of
$X\rightarrow \eta 2\pi$.
  Most likely,  $G$
and $\eta$ could combine to form  the meson $\eta'$, in which case
the process (\ref{end1}) becomes
\begin{eqnarray*}
X\rightarrow 2\pi (\eta G)=2\pi \eta'
\end{eqnarray*}
\vskip-0.45in
$$\hskip0.95in |$$ \vskip-0.53in \begin{equation}\label{end} \hskip1.44in \longrightarrow
\eta \pi\pi \end{equation} Eq.(\ref{end}) is just $(X\rightarrow
\eta 4\pi)$, where the factor of $\eta'\rightarrow \eta \pi\pi$ is
the dominate channel to be considered (i.e., $B(\eta'\rightarrow
\eta \pi\pi)\simeq 65\%$). In this view, the process of
($X\rightarrow \eta 2\pi$) will be almost forbidden and the
($X\rightarrow \eta 4\pi$) or ($X\rightarrow \eta' 2\pi$) would be
 most favorable, i.e.,
\begin{eqnarray}\label{end2}
B(X\rightarrow \eta'2\pi)>> B(X\rightarrow \eta2\pi).
\end{eqnarray}
Since $m_{\eta'}>>m_\eta$, this is a very unusual result. {\it
 This result comes from that $X(1835)$ is a
baryon-antibaryon bound state, and its decay is caused by
$N-\bar{N}$ annihilations.}

\section{Summary}
\noindent In this talk,  the main contents of Refs.\cite{Yan1,Yan2}
 have been revived with some commentary. The main conclusion
 of \cite{Yan1,Yan2} is that
$X(1835)$ can be thought as a baryonium described by the Skyrme
model. There are two positive evidences: 1) There exist a classical
$N\bar{N}$-Skyrmion solution with about $\sim 10$MeV binding
energies in the model; 2) The decay of this Skymion-baryonium is
caused by annihilation of $p-\bar{p}$ inside $X(1835)$ through the
quantum tunneling effect, and hence the most favorable decay
channels are $X\rightarrow \eta 4\pi$ or $X\rightarrow \eta' 2\pi$.
These lead to  reasonable interpreting  the data of BES, and
especially to useful prediction on the decay mode of $X(1835)$ for
the experiment.

It would be interesting to pursue the the semi-classical quantum
corrections in the further works.

\begin{center}
{\bf ACKNOWLEDGMENTS}
\end{center}
I would like to acknowledge my collaborators Professor Bo-qiang Ma,
and Gui-Jun Ding, Si Li and Bin Wu for their great contributions to
the works reviewed in this talk.  We would like also to thank
Professor  Shan Jin and Professor Zhi-Peng Zheng for their
stimulating discussions. This work is partially supported by
National Natural Science Foundation of China under Grant Numbers
9043021, and by the PhD Program Funds of the Education Ministry of
China under Grant Numbers 20020358040, and KJCX2-SW-N10 of the
Chinese Academy of Science.


\end{document}